\documentclass[12pt]{article}
\usepackage{graphicx}
\input epsf
\usepackage{amssymb}
\parskip=\medskipamount
\def\om{\omega}   
\def\ID{\relax{\rm l\kern-.18 em D}}
\def\IE{\relax{\rm l\kern-.18 em E}}
\def\IK{\relax{\rm l\kern-.18 em K}}
\def\IL{\relax{\rm I\kern-.18 em L}}
\def\IN{\relax{\rm I\kern-.18 em N}}
\def\IR{\relax{\rm I\kern-.18 em R}}
\def\uno{\relax{\rm 1\kern-.18 em l}}

\def\IK{\relax{\rm l\kern-.18 em K}}
\def\IL{\relax{\rm I\kern-.18 em L}}
\def\IN{{\Bbb N}}
\def\IR{{\Bbb  R}}
\def\ii{\rm i\,}
\def\Re{\mathop{\rm Re}\nolimits}
\def\Im{\mathop{\rm Im}\nolimits}

\def\smallonehalf{\frac{{}_1}{{}^2}}

\def\wt{\widetilde}
\def\frac#1#2{{#1\over #2}}
\def\fracpd#1#2{\frac{\partial #1}{\partial #2}}

\def\ptos{\leaders\hbox to 2mm{\hfil{.}\hfil}\hfill}
\def\\{\hfill\break}

\def\<#1>{\langle#1\rangle}
\def\ii{{\rm i\,}}

\font\tenfrak=eufm10  \font\sevenfrak=eufm7  \font\fivefrak=eufm5
\newfam\frakfam
\textfont\frakfam=\tenfrak\scriptfont\frakfam=\sevenfrak
\scriptscriptfont\frakfam=\fivefrak

\font\tengoth=eufm10 scaled\magstep1 \font\sevengoth=eufm7
\font\fivegoth=eufm5
\newfam\gothfam
\textfont\gothfam=\tengoth\scriptfont\gothfam=\sevengoth
   \scriptscriptfont\gothfam=\fivegoth
\newtheorem{proposicion}{Proposition}

\begin{document}

\title{A new proof of the higher-order superintegrability  of \\
a noncentral oscillator with inversely quadratic nonlinearities }

\author{ Manuel F. Ra\~nada$^{1}$, Miguel A. Rodr\'{\i}guez$^{2}$ and
 Mariano Santander$^{3}$  \\ [3pt]
$^{1}$ {\sl Dept. de F\'{\i}sica Te\'orica and IUMA, Universidad de Zaragoza} \\
  {\sl 50009 Zaragoza, Spain}  \\ [2pt]
$^{2}$  {\sl Dept.  de F\'{\i}sica Te\'orica, Universidad Complutense de Madrid} \\
  {\sl 28040 Madrid, Spain}  \\ [2pt]
$^{3}$   {\sl Dept.  de F\'{\i}sica Te\'orica, Universidad de Valladolid} \\
  {\sl 47011 Valladolid, Spain}   }
 \date{February 19, 2010}
%-------------------------------------------------------------------
%%  Version revisada no. 2
%%  Nuevo parrafo en las pag. 12-13
%%  Enviada a JMP el 19 de Febrero de 2010
%-------------------------------------------------------------------
\maketitle

\begin{abstract}
The superintegrability of a rational harmonic oscillator
(non-central harmonic oscillator with rational ratio of
frequencies) with non-linear ``centrifugal" terms is studied. In
the first part, the system is directly studied in the Euclidean
plane; the existence of higher-order superintegrability (integrals
of motion of higher order than 2 in the momenta)  is proved by
introducing a deformation in the quadratic complex equation of the
linear system. The constants of motion of the nonlinear system are
explicitly obtained. In the second part, the inverse problem is
analyzed in the general case of $n$ degrees of freedom; starting
with a general Hamiltonian $H$, and introducing appropriate
conditions for obtaining superintegrability,  the particular
``centrifugal" nonlinearities   are obtained.
\end{abstract}

\begin{quote}
%---------------
Harmonic oscillators. Nonlinear systems.

{\sl Running title:}{\enskip}
Higher-order superintegrability of a nonlinear ocillator.

%---------------
%%  37J35    Completely integrable systems, topological structure
%%  70H06   Completely integrable systems and methods of integration

%---------------
%%  02.30.Ik    Integrable systems
%%  05.45.-a   Nonlinear dynamics and chaos
\end{quote}

%---------------
\footnoterule{\small
\begin{quote}
{\tt E-mail: {mfran@unizar.es}; {rodrigue@fis.ucm.es}; {msn@fta.uva.es};  }
\end{quote}
}

%---------------
\newpage
%---------------

%-------------------------------------------------------------------
%%  Section 1
\section{Introduction}

 A superintegrable system is a system that is integrable
(in the sense of Liouville-Arnold) and that, in addition to this,
possesses more constants of motion than degrees of freedom.  At
this point we must note that the maximum number $N$ of functionally
independent constants of motion for a system in a $d$-dimensional 
manifold is $N=d-1$.
Thus if a Lagrangian (or Hamiltonian) system has $n$ degrees of 
freedom  then, as the phase space is $2n$-dimensional, we have 
that the maximum  number of independent constants of motion is $
N = 2 n - 1$.  There are three well known examples of this very 
particular class of systems, namely, the Kepler problem, the 
isotropic harmonic oscillator, and the non-isotropic oscillator 
with commensurable frequencies.  The
two-dimensional harmonic oscillator is a system trivially
integrable since it can be considered as a kind of ``direct sum"
of two systems with one degree of freedom. If the oscillator is
isotropic then it has the angular momentum as an additional
integral of motion. If the oscillator is non-isotropic the angular
momentum is not preserved as the potential is not central;
nevertheless when the quotient of the two frequencies is a
rational number then the system has another additional integral.
Concerning the three-dimensional Kepler problem, it possesses not
only the energy and the angular momentum as constants of motion,
but also the Runge-Lenz vector; only five of these seven integrals
are functionally independent since in this case the number of
degrees of freedom is $n=3$.   In these three cases it is well
known that all the orbits became closed for the case of bounded
motions. This high degree of regularity (existence of periodic
motions) is a consequence of the superintegrable character.

Fris et {\sl al} \cite{FrMS65} studied the  two-dimensional
Euclidean systems which admit separability in two different
coordinate systems and obtained four families of  potentials
$V_r$, $r=a,b,c,d$,  possessing three functionally independent
integrals of the motion. A very important point is that these
three constants of motion were linear or quadratic in the
velocities (momenta). In fact, if we call superseparable a system
that admits Hamilton-Jacobi  separation of variables
(Schr\"odinger in the quantum case) in more than one coordinate
system, then quadratic superintegrability (i.e.,
superintegrability with linear or quadratic constants of motion)
can be considered as a property arising from superseparability.
The first two families
%---------------
\begin{eqnarray*}
 V_a  &=& {\smallonehalf}\om_0^2(x^2 + y^2)
 + {k_2\over x^2}  + {k_3\over y^2} \,,\cr
 V_b  &=&  {\smallonehalf}\om_0^2(4 x^2 + y^2)
 + k_2 x + {k_3\over y^2}     \,,
\end{eqnarray*}
%---------------
can be considered as the more general deformations (with strengths
$k_2$, $k_3$) of the 1:1 and 2:1 harmonic oscillators preserving
quadratic superintegrability (the other two families, $V_c$ and
$V_d$, were related with the Kepler problem). The
superintegrability of $V_a$ was later on studied by Evans
\cite{Ev90,Ev91} in the more general case of $n$ degrees of
freedom.

 A natural generalization of $V_a$ is given by the following potential
$$
  V_a(\om_1,\om_2) =  {\smallonehalf}({\om_1}^2 x^2 + {\om_2}^2 y^2)
        + \frac{k_1}{x^2}  + \frac{k_2}{y^2} \,,
$$
that contains a more general  harmonic oscillator with anisotropy.
This new potential is only separable in Cartesian coordinates and
its superintegrability (if it exists) must be of higher-order
('higher-order superintegrability' means that some of the
integrals of motion are polynomials in the momenta of order higher
than 2). Therefore, the method of the multiple separability cannot
be used and it must be studied by making use of a different
approach. In fact, the superintegrability of this nonlinear
system, for the case of rational ratio of the frequencies, was
first proved by Ra\~nada et {\sl al} in \cite{RaSaMontreal} using
the properties of the isotonic (or singular) oscillator and the
Pinney-Ermakov equation \cite{Pi50}  for obtaining a complex
factorization. More recently  Evans et {\sl al} \cite{EvVe08b} and
Rodr\'{\i}guez et {\sl al} \cite{RodTW08,RodTW09} have also proved
this property using  the geometric formalism of dimensional reduction.
It have been proved that certain nonlinear integrable systems
may arise as reductions of very simple systems defined in
higher-dimensional spaces (see e.g. \cite{GrLanMar94, CaClMa07}).
In this particular case the authors start with an harmonic
oscillator in a higher-dimensional space and then the dimensional
reduction introduces the nonlinearities but in a way that
preserves the superintegrability. We note that this geometric
method has been also used for proving the superintegrability of
the Kepler--Coulomb system with nonlinear terms \cite{RodTW09,
EvVe08a, BaHeJpa09}.

 Now,  we present a new method to prove the higher-order
superintegrability of this nonlinear system. This new method, that
is more straightforward than the previously known methods, is
directly  related with the approach presented in
\cite{RaSaMontreal} (but without making use of the properties of
the Pinney-Ermakov equation) and, at the same time, it is also
related with some of the results obtained in
\cite{RodTW08,RodTW09}.

 The plan of the article is as follows:
in Sec. 2 we present the problem from the Lagrangian viewpoint and
we study the superintegrability of the two dimensional potential
$V_a(\om_1,\om_2)$ using the existence of a complex factorization
as an approach.  First we consider the linear harmonic oscillator
and then we prove the superintegrability of the nonlinear system.
The constants of motion of the nonlinear system are explicitly
obtained.   In Sec. 3 we analyze the general case of  $n$ degrees
of freedom. It is written by using the  Hamiltonian approach and
it has a more generic character. We start with a very general
Hamiltonian $H$  and then we introduce the restrictions to
obtain superintegrability (but without making use of the
property of separability). The idea is that this more general
approach could be used in future papers as a starting point for
the search of other more general superintegrable systems. Finally
in Sec. 4 we make some comments and present some open questions.

%-------------------------------------------------------------------
%%  Section 2
\section{Complex factorization and superintegrability }

%---------------
\subsection{Superintegrability of the linear Harmonic Oscillator }\label{2.1}

The two-dimensional  harmonic oscillator
$$
  L_{HO} =  {\smallonehalf} (v_x^2 + v_y^2)
  -   {\smallonehalf} ({\om_1}^2 x^2 + {\om_2}^2 y^2)
$$
has the two partial (one-degree of freedom) energies, $I_1=E_x$ and
$I_2=E_y$, as fundamental integrals.
The superintegrability of the rational case,
$\om_1 =n_x {\om_0}$, $\om_2 = n_y {\om_0}$,
with integers $n_x, n_y$, can be proved by making use
of a complex formalism \cite{JauHillPr40,Perelomov,RaSaJmp02}.
Let $K_i$, $i=x,y$, be the following two complex functions
$$
 K_x = v_x + {\ii} n_x {\om_0}\, x \,,{\quad}
 K_y = v_y + {\ii} n_y {\om_0}\, y \,,
$$
then we have the following time-evolution
$$
  \frac{d}{dt}K_x = {\ii} n_x {\om_0}\,K_x \,,{\quad}
  \frac{d}{dt}K_y = {\ii} n_y {\om_0}\,K_y \,.
$$
Thus, the functions $K_{ij}$ defined as
$$
 K_{ij} = (K_i)^{n_y} (K_j^{*})^{n_x} \,,{\quad} i,j=x,y,
$$
are  constants of motion. The two real functions $|K_{xx}|^2$ and
$|K_{yy}|^2$ are proportional to the energies $E_x$ and $E_y$ and
concerning $K_{xy}$, since it is a complex function, it determines
not one but two real first integrals, $\Im (K_{xy})$ and $\Re
(K_{xy})$. So, we have obtained four integrals but, since the system
is two-dimensional, only three of them can be independent. We can
choose $I_1=E_x$, $I_2=E_y$, and $I_3=\Im (K_{xy})$ as the set of
fundamental constants of motion (the other constant $I_4=\Re
(K_{xy})$ can be expressed as a function of $E_x$, $E_y$, and $\Im
(K_{xy})$).   

 As an example, for the Isotropic case, $\om_1=\om_2=\om_0$,
we obtain
%---------------
\begin{eqnarray*}
  I_3  &{\equiv}&   \frac{1}{\om_0}\Im(K_{xy}) = x v_y - y v_x   \,,\cr
  I_4  &{\equiv}&   \Re(K_{xy}) = v_x v_y + {\om_0}^2 x y  \,,
\end{eqnarray*}
($I_3$ is the angular momentum and $I_4$ the component $F_{xy}$ of
the Fradkin tensor \cite{Frad65}) and for the first non-isotropic
case, $\om_1 = 2 \om_0$, $\om_2 = \om_0$, we arrive to
%---------------
\begin{eqnarray*}
  I_3  &{\equiv}&  \frac{1}{2\om_0} \Im(K_{xy}) = (x v_y - y v_x) v_y
  - {\om_0}^2 x y^2  \,,\cr
  I_4  &{\equiv}&  \Re(K_{xy}) =  v_x v_y^2
  + {\om_0}^2 (4 x v_y - y v_x)\,y  \,.
\end{eqnarray*}
The integral $I_3$ of the 3:1 oscillator will be cubic and, in the
general $n_x\!:\!n_y$ case, the function $I_3$ will be a
polynomial in the velocities (momenta) of degree $n_x+n_y-1$.

%---------------
\subsection{Superintegrability of the nonlinear system }

The technique presented in the previous section proves not only
the super-integrability of the rational case but also the
existence of a complex factorization for the additional constant
of motion. Now we assume, as starting point of our approach, that
if a new system can be obtained by a deformation of the harmonic
oscillator then it must be also endowed with a similar property.

The analysis will proceed in two steps.

%---------------
\noindent{\sl  Step 1.} {\sl The associated quadratic equation }

Let $K=a + {\ii} b$ be such that $dK/dt =  {\ii} n {\om} K$. Then the
function $K_2$ defined as $K_2=K^2=(a^2-b^2)+2 {\ii} ab$ satisfies
$dK_2/dt = 2 {\ii} n {\om} K_2$. Thus the time evolution of the
functions
$$
 K_{2x}=(v_x^2 - n_x^2 \om_0^2 x^2) + 2\,{\ii} n_x {\om_0} x v_x\,,{\quad}
 K_{2y}=(v_y^2 - n_y^2 \om_0^2 y^2) + 2\,{\ii} n_y {\om_0} y v_y\,,
$$
  is given by
$$
  \frac{d}{d t}\,K_{2x} = 2\,{\ii} n_x {\om_0} K_{2x}  \,,{\qquad}
  \frac{d}{d t}\,K_{2y} = 2\,{\ii} n_y {\om_0} K_{2y}  \,,
$$
and hence the complex functions $K_{2ij}$ defined as
$$
  K_{2ij} = (K_{2i})^{n_j}\,(K_{2j}^{*})^{n_i} \,,{\quad} i,j=x,y,
$$
are constants of motion.

 The two complex functions, $K_{ij}$ and $K_{2ij}$, must be
considered as two alternative ways to prove superintegrability
but, of course, the first one is simpler than the quadratic. As an
example we have $ |K_{2xx}|= |K_{xx}|^2$ and $ |K_{2yy}|=
|K_{yy}|^2$. In the general $n_x\!:\!n_y$ case, the functions $I_3
= \Im (K_{2xy})$ and $I_4 = \Re (K_{2xy})$ will be a polynomials
in the velocities (momenta) of degree $2(n_x+n_y-1)-1$ and
$2(n_x+n_y)$. So, if the  study is restricted to the harmonic
oscillator,  it is better to use the function $K_{ij}$ since it leads
to simpler expressions for the constants of the motion.

%---------------
\noindent{\sl Step 2.} {\sl  Introducing a deformation in the
``quadratic equation" }

Let us now consider the following $(F,G)$-dependent family of
potentials
$$
 V(n_x,n_y,F,G) = {\smallonehalf}{\om_0}^2(n_x^2 x^2 + n_y^2 y^2)
 + {\smallonehalf} F(x) + {\smallonehalf}G(y) \,.
$$
The problem is to determine which particular values of the
functions $F$ and $G$ can preserve the existence of a complex
factorization.  At this point we assume that, in order to solve
this problem, it is more convenient to use the quadratic equation
since it seems as more 'deformable' than the linear one.

Let us denote by $A_j$ and $B_j$, $j=x,y$, the following functions
%---------------
\begin{eqnarray*}
  A_x &=& v_x^2 - n_x^2 \om_0^2 x^2 + F(x) \,,{\quad}
  B_x = 2  n_x {\om_0} x  v_x   \,,    \cr
  A_y &=& v_y^2 - n_y^2 \om_0^2 y^2 + G(y) \,,{\quad}
  B_y = 2 n_y {\om_0} y  v_y    \,.
\end{eqnarray*}
%---------------
Then, if we require for the functions $A_j$ and $B_j$ a time-evolution
of the form
$$
 \frac{d}{dt}\,A_j = -\,2 n_j {\om_0}\,B_j  \,,{\quad}
 \frac{d}{dt}\,B_j =    2 n_j {\om_0}\, A_j \,,{\quad} j=x,y \,,
$$
we arrive (we omit the details) to the first-order differential
equations
$$
 x F' + 2 F = 0  \,,{\quad}  y G' + 2 G = 0\,,
$$
with solutions $F=k_1/x^2$ and $G=k_2/y^2$ with arbitrary
constants $k_1$ and $k_2$. Therefore, only if $F$ and $G$ have
this particular structure, the complex functions $M_j$ defined as
$M_j=A_j+{\ii}B_j$ play,  in this nonlinear case, a similar role
to the complex functions  $K_j$ of the linear case.

%---------------
%%  (Proposicion 1)
\begin{proposicion}
Consider the non-linear potential
$$
  V_a(n_x,n_y)  =  {\smallonehalf}{\om_0}^2(n_x^2 x^2 + n_y^2 y^2)
    + \frac{k_1}{2x^2}  + \frac{k_2}{2y^2}    \label{Van1n2}
$$
representing an harmonic oscillator with rational ratio of
frequencies, ${\om_1}= n_x{\om_0}$, ${\om_2}= n_y{\om_0}$, and
inversely quadratic nonlinearities and let  us denote by $M_j$,
$j=x,y$, the following two complex functions
$$
 M_{x} = \Bigl(v_x^2 - n_x^2 \om_0^2 x^2 +\frac{k_1}{x^2}\Bigr) + 2\,{\ii} n_x {\om_0} x v_x\,,{\quad}
 M_{y} = \Bigl(v_y^2 - n_y^2 \om_0^2 y^2 +\frac{k_2}{y^2}\Bigr) + 2\,{\ii} n_y {\om_0} y v_y\,.
$$
Then, the complex functions $M_{ij}$ defined as
$$
  M_{ij} = (M_i)^{n_j}\,(M_j^{*})^{n_i}\,,{\quad} i,j=x,y,
$$
are constants of the motion.
\end{proposicion}
Firstly, the moduli of $M_x$ and $M_y$ are given by
$$
 |M_x|^2 = 4 (E_x^2 - k_1 n_x^2 \om_0^2) \,,{\quad}
 |M_y|^2 = 4 (E_y^2 - k_y n_y^2 \om_0^2) \,.
$$
The time-evolution of the functions $M_x$ and $M_y$ is given by
$$
 \frac{d}{d t}\,M_x = 2\,{\ii} n_x {\om_0}\,M_x  \,,{\quad}
 \frac{d}{d t}\,M_y = 2\,{\ii} n_y {\om_0}\,M_y  \,.
$$
Thus we have
%---------------
\begin{eqnarray*}
  \frac{d}{dt}\,M_{xy} &=&  n_y (M_x)^{(n_y-1)}\,(M_y^{*})^{n_x}\dot{M_x}
   +  n_x (M_x)^{n_y}\,(M_y^{*})^{(n_x-1)}\dot{M_y}^{*}   \cr
  &=& (2{\ii} \om_0) \bigl( n_y n_x - n_x n_y \bigr)
  (M_x)^{n_y}\,(M_y^{*})^{n_x} =  0  \,.
\end{eqnarray*}
As in the linear case, $M_{xy}$ can be considered as
coupling the two degrees of freedom.

Hence the potential $V_a(n_x,n_y)$ is superintegrable for any
rational  value of the the quotient $\om_2/\om_1$. In the
particular 1:1 case, $\om_1=\om_2=\om_0$, the potential reduces to
the $V_a$ potential
$$
  V_a(1,1) \equiv V_a =  {\smallonehalf} {\om_0}^2 (x^2 + y^2)
  +  \frac{k_1}{2x^2} + \frac{k_2}{2y^2}
$$
(that is the only superseparable potential in this family) and in
the general case, $\om_1 = n_x\,\om_0$, $\om_2 = n_y\,\om_0$,  the
potential $V_a(n_x,n_y)$  represents a generalized $V_a$ potential
with a non-isotropic  $n_x\!:\!n_y$ oscillator (note that in the
2:1 case, the potential $V_a(2,1) $ {\sl is different} from the family
$V_b$).

In the general $n_x\!:\!n_y$ case, the functions $ \Re(M_{xy})$
and $\Im(M_{xy})$ will be polynomials  in the velocities (momenta)
of degree $2(n_x+n_y)$ and  $2(n_x+n_y)-1$ respectively. The real
part takes the form
$$
 \Re(M_{xy})  =  2^{(n_x+n_y)} (E_x)^{n_y} (E_y)^{n_x} +
 {\lambda}\om_0^2 J_3  \,,
$$
where $\lambda$ is a numerical coefficient.  Thus the additional
constant of $V_a(n_x,n_y)$ is in fact the function $J_3$ that is
of degree $2(n_x+n_y-1)$. If we denote  by $I_3$ the constant  of
motion of the associated linear system (of degree $n_x+n_y-1$),
then  the additional third integral  $J_3$ can be written as
follows
$$
  J_3 = I_3^2  + k_1 J_3^{(10)} + k_2 J_3^{(01)}+ k_1^2 J_3^{(20)}
        + k_1 k_2 J_3^{(1,1)} + \dots +  k_2^{n_x} J_3^{(0\,n_x)}\,.
$$
That is, the integral $J_3$ of the non-linear system ($k_1\neq 0$,
$k_2\neq 0$) appears as a deformation, not of the function $I_3$
itself, but of its square $I_3^2$ (this property was already
mentioned in \cite{RaSaMontreal}).

Next, we give the expressions of the constant $J_3$ for the three
first cases:
\begin{enumerate}
\item[(i)] Potential $V_a(1,1)$ corresponding to a central
(isotropic) harmonic oscillator.

In this case we have $\om_1 = \om_2 = \om_0$ and the integral of
motion $I_3$ of the associated linear system is just the angular
momentum, $I_3 = (1/\om_0)\Im(K_{xy})  =  x v_y - y v_x$.  Then we
have
$$
 \Re(M_{xy})  =  4 E_x E_y - 2  \om_0^2 J_3  \,,
$$
with $J_3$ given by
$$
 J_3 = I_3^2  +  k_1 \Bigl(\frac{y}{x}\Bigr)^2
 +  k_2 \Bigl(\frac{x}{y}\Bigr)^2\,.
$$

\item[(ii)] Potential $V_a(2,1)$ corresponding to a non-isotropic
2:1 oscillator with frequencies $\om_1 = 2\om_0$, $\om_2 = \om_0$.

If we denote by $I_3$ the integral of motion of the associated
linear system
$$
   I_3 = \frac{1}{2\om_0}\Im(K_{xy})  =  (x v_y - y v_x) v_y - \om_0^2 x y^2 \,,
$$
then we obtain
$$
 \Re(M_{xy})  =  8 E_x E_y^2 - 8 \om_0^2 J_3  \,,
$$
with $J_3$ given by
$$
 J_3 = I_3^2  +  k_1\Bigl(\frac{y^2}{x^2}\Bigr)v_y^2 +
 \frac{k_2}{2y^2}(yv_x-2xv_y)^2 + \frac{k_1k_2}{2x^2} +
 k_2^2\Bigl(\frac{x^2}{y^4}\Bigr) \,.
$$

\item[(iii)] Potential $V_a(3,1)$ corresponding to a non-isotropic
3:1 oscillator with frequencies $\om_1 = 3\om_0$, $\om_2 = \om_0$.

If we denote by $I_3$ the constant of motion of the associated
linear system
$$
  I_3 = \frac{1}{\om_0}\Im(K_{xy}) = 3 (x v_y - y v_x) v_y^2
  + \om_0^2 (y v_x - 9 x v_y) y^2  \,,
$$
then we obtain
$$
 \Re(M_{xy})  =  16 E_x E_y^3 - 2 \om_0^2 J_3   \,,
$$
with $J_3$ given by
$$
  J_3 = I_3^2 + k_1\,J_3^{(10)}  + k_2\,J_3^{(01)}
 + k_1k_2\,J_3^{(1,1)} +  k_2^2\,J_3^{(0,2)} + k_1k_2^2\,J_3^{(1,2)}
 + k_2^3\,J_3^{(0,3)}\,,
$$
with the functions $J_3^{(10)}$, $J_3^{(01)}$, $J_3^{(1,1)}$,
$J_3^{(0,2)}$ and $J_3^{(0,3)}$,  given by
%---------------
$$\begin{array}{cclrcl}
 J_3^{(10)}  &=& \displaystyle
  \frac{y^2}{x^2}\,(3 v_y^2- \om_0^2 y^2)^2   \,, &
 J_3^{(01)}  &=& \displaystyle
 \frac{3}{y^2}\,( 2 y v_x v_y - 3 x v_y^2 + 3\om_0^2 x y^2)^2   \,, \cr
  J_3^{(1,1)} &=&  \displaystyle  \frac{12 v_y^2}{x^2}   \,, &
  J_3^{(0,2)} &=&
   \displaystyle  \frac{3}{y^4}(3 x v_y - y v_x)^2  \,,\cr
  J_3^{(1,2)} &=& \displaystyle  \frac{3}{x^2y^2}   \,,  &
  J_3^{(0,3)} &=& \displaystyle  \frac{9x^2}{y^6}   \,.
\nonumber
\end{array}
$$
%---------------
\end{enumerate}

Summarizing, within the $V_a(n_x, n_y)$ family, only in the
particular isotropic 1:1 case the function $J_3$ is quadratic in
the velocities and, because of this, only in this case the
superintegrability arises from separability in two different
coordinate systems. In all the remaining cases, $V_a(n_x,n_y)$ is
a superintegrable but not superseparable potential.

%-------------------------------------------------------------------
%%  Section 3
\section{Hamiltonian formalism and $n$ degrees of freedom }

The previous section was directly focused on  the two dimensional
potential $V_a(n_x,n_y)$.    Now we present the study of the
general case of $n$ degrees of freedom but using a rather
different approach. The idea is to start with an integrable but
very general Hamiltonian $H$ and then look for the properties to
be satisfied by $H$ in order to admit  superintegrability
determined by appropriate complex functions $K_i$.

Let $H$ be the following Hamiltonian
$$
 H = {\smallonehalf} \sum_{i=1}^{n} p_i^2 + V
 \,,{\quad} V = \sum_{i=1}^{n} V_i(x_i)  \,,
$$
defined in a $2n$-dimensional phase space $T^*Q$ ($Q$ is the
$n$-dimensional configuration space)   endowed with the standard
Poisson bracket
$$
 \bigl\{R,S\bigr\} = \sum_i\,\biggl(\fracpd{R}{x_i}\,\fracpd{S}{p_i}
 - \fracpd{R}{p_i}\,\fracpd{S}{x_i}\biggr)  \,.
$$
It is clear that $H$ is integrable; now we set out the problem to
determine the expressions of the functions $V_i(x_i)$ to admit
superintegrability (but without making use of the property of
separability).

Let us now define a set of $n$ functions (possibly complex) linear
in the momenta
$$
 K_i = p_i + f_i(x_i)  \,,{\quad} i=1,2,\dots,n,
$$
(note that we have chosen the coefficient of $p_i$ equal to 1).
Then, it is evident that
$$
 \{ K_i\,, K_j\} = 0  \,,
$$
but
$$
 \{ K_i\,, H\} =  f_i'(x_i) p_i - V_i'(x_i) \ne 0 \,,
$$
(except in the trivial case $f_i(x_i)={\rm constant}$ and
$V_i(x_i) = {\rm constant}$). Our next step is to write the
Hamiltonian in terms of the $K_i$ functions.  Since we assume that
these functions can be complex, we impose
$$
 H= {\smallonehalf} \sum_{i=1}^nK_iK_i^*= {\smallonehalf} \sum_{i=1}^n|K_i|^2
$$
where $K_i^*$ is the complex conjugate of $K_i$. So we have
$$
  |K_i|^2 = p_i^2+p_i\bigl(f(x_i)+f^*(x_i)\bigr)+|f(x_i)|^2 \,.
$$
If we assume that the Hamiltonian $H$ is quadratic in the momenta,
and without linear terms, then the functions $f_i(x_i)$ should be
pure imaginary. So we arrive to
$$
 K_i = p_i + {\ii} g_i(x_i) \,,{\quad}
 V = {\smallonehalf} \sum_{i=1}^n g_i(x_i)^2 \,,
$$
with $g_i(x_i)$ real functions.

%---------------
\subsection{From integrability to superintegrability via the functions $K_i$ }

The Hamiltonian $H$ is trivially integrable, the $n$ functions
$|K_i|^2$ are functionally independent constants of motion and the
Hamiltonian is half their sum. In order to study its
superintegrability, we make use of  the tensor $T_{ij}$ defined by
$$
 T_{ij}=K_iK_j^* {\quad}{\rm  with}{\quad}
 H = {\smallonehalf}\sum_{i=1}^n T_{ii}  \,.
$$
Since the Poisson bracket of the functions $K_i$ with the
Hamiltonian $H$ is now given by
$$
\{K_i,H\}=  g_i'(x_i)\bigl( {\ii} p_i -  g_i(x_i)\bigr)= {\ii} g_i'(x_i)K_i \,,
$$
then the off diagonal tensor components have the following Poisson
bracket with the Hamiltonian
$$
 \{T_{ij},H\} = \{K_i,H\}K_j^*+K_i\{K_j^*,H\} =
 {\ii} \big(g_i'(x_i)-g_j'(x_j)\big)T_{ij} \,.
$$
Thus  if the $g_i'(x_i)$ are a numerical constant (independent of
$i$) then the functions $T_{ij}$ are constants of the motion.
This happens with
$$
 g_i(x_i) = {\om_0} x_i\,,{\quad} i=1,\ldots, n,
$$
so we get the isotropic harmonic oscillator in $n$ dimensions
which is certainly superintegrable:
$$
 H_1 = {\smallonehalf} \sum_{i=1}^n p_i^2
 + {\smallonehalf} {\om_0}^2\sum_{i=1}^n x_i^2\,.
$$

%---------------
\subsection{Generalizing the Hamiltonian $H_1$}

If we consider, instead of the tensor $T$, a new tensor  $\wt{T}$
with components
$$
 \wt{T}_{ij} = K_i^{n_j}(K_j^*)^{n_i} \,,{\quad}  i,j=1,\ldots,n,
$$
(where $n_i$ are positive integers) we can repeat the arguments
above and obtain
%---------------
\begin{eqnarray*}
\{\wt{T}_{ij},H\} &=&
n_j\{K_i,H\}K_i^{n_j-1}(K_j^*)^{n_i}+n_iK_i^{n_i}\{K_j^*,H\}(K_j^*)^{n_i-1}\cr
&=& {\ii} \big(n_jg_i'(x_i)-n_ig_j'(x_j)\big)\wt{T}_{ij}    \,.
\end{eqnarray*}
Thus  if the functions $g_i(x_i)$ satisfy the relations
$$
 n_jg_i'(x_i)-n_ig_j'(x_j) = 0 \,,{\quad}  i,j=1,\ldots,n,
$$
then the functions $\wt{T}_{ij}$ (components of the new tensor
$\wt{T}$) are constants of the motion. In this way we obtain the
Hamiltonian of the rational anisotropic harmonic oscillator
(non-central harmonic oscillator with rational ratio of
frequencies)
$$
 H_2 = {\smallonehalf}\sum_{i=1}^n p_i^2
 + {\smallonehalf}{\om_0}^2\sum_{i=1}^n n_i^2x_i^2
$$
which represents the most simple superintegrable generalization of
the Hamiltonian $H_1$.

%---------------
\subsection{Generalizing the Hamiltonian $H_2$}

In order to obtain a superintegrable generalization of the
Hamiltonian $H_2$, we first consider the squares of the functions $K_i$
$$
 K_i^2 = p_i^2-n_i^2 {\om_0}^2 x_i^2 + 2\,{\ii}n_i{\om_0} x_ip_i
 \,,{\quad} i=1,\ldots n,
$$
satisfying
$$
 \{K_i^2,H_2\} = \{K_i,H_2\}K_i + K_i\{K_i,H_2\} = 2\,{\ii} n_i{\om_0} K_i^2 \,,
$$
and then we introduce a deformation of  the real part
$$
 M_i = p_i^2-n_i^2{\om_0}^2 x_i^2+h_i(x_i) + 2\,{\ii} n_i{\om_0} x_ip_i
 \,,{\quad} i=1,\ldots n.
$$
Now let us consider the following Hamiltonian
$$
 H = {\smallonehalf}\sum_{i=1}^n p_i^2 + {\smallonehalf}{\om_0}^2\sum_{i=1}^n n_i^2x_i^2 + \sum_{i=1}^nV_i(x_i) \,,
$$
where $V_i(x_i)$ are some functions to be determined by imposing
that $|M_i|^2$ Poisson commutes with $H$. We have
$$
\{|M_i|^2,H\} =  2 \bigl(h_i'-2 V_i'\bigr)\, p_i^3 + 2 \Bigl[ 2 \bigl(h_i'-2 V_i'\bigr)h_i - n_i^2 {\om_0}^2 x_i^2  \bigl(4 h_i +x_i h_i' + 2 x_iV'_i\big)  \Bigr]\, p_i   \,,
$$
so that we arrive at
$$
h_i'-2 V_i'=0 \,,{\quad} 4 h_i +x_i h_i' + 2 x_iV'_i=0 \,,
$$
with solution
$$
V_i(x_i)=\frac{1}{2}h_i(x_i)=\frac{k_i}{2x_i^2},\quad i=1,\ldots , n,
$$
(up to inessential additive constants).
Hence, the above Hamiltonian $H$ becomes
$$
 H_3 = {\smallonehalf}\sum_{i=1}^n p_i^2 + {\smallonehalf}{\om_0}^2\sum_{i=1}^n n_i^2x_i^2 + {\smallonehalf}\sum_{i=1}^n\frac{k_i}{x_i^2}  \,,
$$
in such a way that the $n$ functions  $|M_i|^2$ coincide with the
square of the $n$ energies $E_i $ up to a constant
$$
 |M_i|^2 = 4 (E_i^2 - k_i n_i^2 {\om_0}^2) \,,{\quad}
  E_i = \frac{1}{2} p_i^2+\frac{1}{2}{\om_0}^2  n_i^2x_i^2+  \frac{k_i}{2x_i^2}  \,.
$$
The important point is  that the functions $M_i$ satisfy
$$
 \{M_i,H_3\} = 2\,{\ii} n_i {\om_0}\,M_i \,.
$$
Hence, if we denote by  $M_{ij}$ the functions defined by the
products
%---------------
\begin{eqnarray*}
 M_{ij}  &=&  M_i^{n_j}(M_j^*)^{n_i} =
\Bigr(p_i^2-n_i^2{\om_0}^2 x_i^2+\frac{k_i}{x_i^2}+2\,{\ii} n_i{\om_0} x_ip_i\Bigl)^{n_j}     \cr
&\times&  \Bigr(p_j^2-n_j^2{\om_0}^2 x_j^2+\frac{k_j}{x_j^2}-2\,{\ii}  n_j{\om_0} x_jp_j\Bigl)^{n_i}  \,,
\end{eqnarray*}
then we have
$$
 \{M_{ij},H_3\} = 0 \,,{\quad} i,j=1,\ldots, n,
$$
what means that both the real part and the imaginary part of
$M_{ij}$ are constants of the motion for $H_3$ (the diagonal
functions $M_{ii}$ are real and, as we have seen, they correspond
to the energies $E_i$). As the system has $n$ degrees of freedom
(and the phase space is $2n$-dimensional) the maximum number of
independent constants of motion is $N = 2 n - 1$; so it is clear
that not all of these quantities will be functionally independent
but, nevertheless, we can extract from them a fundamental set of
$2n-1$ functionally independent invariants.  So we conclude that
the Hamiltonian $H_3$ is superintegrable for all the values of the
integer numbers $n_i$ and for arbitrary values of the constants
$k_i$.  

Next we prove the above statement (it is always possible to extract, 
from the large number of constants of motion, a fundamental set of $2 n - 1$ 
functionally independent functions) by following the same arguments used for 
the harmonic oscillator (isotropic and nonisotropic).  We first recall the following two points 
\begin{itemize}
\item[(i)]     In the isotropic $n$-dimensional case,  the  Fradkin tensor $F$ \cite{Frad65} is represented by a symmetric $n$-dimensional matrix $F_{ij}=p_ip_j+{\om_0}^2 x_ix_j$, $i,j=1,2,\dots,n$, so that  it provides a total set of $(1/2)n (n+1)$ constants of motion. The integrability is consequence of the $n$ diagonal entries $F_{ii}$ (related with the energies $E_i$) that are independent in a trivial way; so we have
$$
  dF_{11}{\wedge}dF_{22}{\wedge}\dots{\wedge}dF_{nn}\ne 0 \,. 
$$
For proving the superintegrability we recall that every nondiagonal function $F_{ij}$ only depends of the four variables $(x_i,p_i, x_j,p_j)$.  Thus we can add, for example,  the $n-1$ entries $F_{jj+1}$ of the upper-next-diagonal so that we obtain 
$$
  dF_{11}{\wedge}dF_{22}{\wedge}\dots{\wedge}dF_{nn}{\wedge}
  dF_{12}{\wedge}dF_{23}{\wedge}\dots{\wedge}dF_{n-1n}\ne 0 \,. 
$$
Hence, these $N = 2 n - 1$ constants of motion are functionally independent.  

\item[(ii)]     In the non-isotropic $n$-dimensional case, the method of the complex factorization \cite{JauHillPr40,Perelomov} discussed in subsection (\ref{2.1}) leads to a complex Hermitian $n$-dimensional matrix $K_{ij}$, $i,j=1,2,\dots,n$, with $K_{ij}$ only depending of the four variables $(x_i,p_i, x_j,p_j)$; so the property of independence 
 $$
  dK_{11}{\wedge}dK_{22}{\wedge}\dots{\wedge}dK_{nn}{\wedge}
  d(\Im{K_{12}}){\wedge} d(\Im{K_{23}}){\wedge}\dots{\wedge} d(\Im{K_{n-1n}})\ne 0 \,, 
$$
is also true in this case (of course in this complex case it is also possible to choose the Real part of the functions $K_{ij}$). 
\end{itemize}

Now, in the case of the nonlinear Hamiltonian $H_3$, we have obtained  a complex Hermitian $n$-dimensional matrix $M_{ij}$, $i,j=1,2,\dots,n$, that can be considered as a nonlinear deformation of the matrix $K_{ij}$.  But, since  $M_{ij}$ only depends of the four variables $(x_i,p_i, x_j,p_j)$, we also have the following property 
$$
  dM_{11}{\wedge}dM_{22}{\wedge}\dots{\wedge}dM_{nn}{\wedge}
  d(\Im{M_{12}}){\wedge} d(\Im{M_{23}}){\wedge}\dots{\wedge} d(\Im{M_{n-1n}})\ne 0 \,. 
$$
Thus, we have proved the existence of a set of $N = 2 n - 1$  functionally independent constants of motion.

We close this section with an interesting property. If we return
(for ease of notation) to  two degrees of freedom, then in the
case of the harmonic oscillator $(k_1= 0, k_2= 0)$, the constant
of the motion $I_4=\Re(K_{xy})$ can be obtained (up to a factor)
as the Poisson bracket of $I_3=\Im(K_{xy})$ with the energy $E_x$
(for example, in the 1:1 case the component $F_{xy}$ of the
Fradkin tensor arises as the Poisson bracket of the angular
momentum $J$ with $E_x$).  This property is preserved by the
deformation $(k_1\ne 0, k_2\ne 0)$ and remains true for the
nonlinear  system.
%---------------
%%  (Proposicion 2)
\begin{proposicion}
The Poisson brackets of $\Re(M_{xy}) $ and $\Im(M_{xy})$ with
$E_x$ are given by
%---------------
\begin{eqnarray*}
  \{\Im(M_{xy})\,,E_x\} &=& 2 \om_0 n_x n_y \Re(M_{xy})  \,,\cr
  \{\Re(M_{xy})\,,E_x\} &=& -\,2 \om_0 n_x n_y \Im(M_{xy})  \,.
\end{eqnarray*}
\end{proposicion}

Note that the second Poisson bracket means that $\Im(M_{xy})$
(degree $2(n_x+n_y)-1$) is just the Poisson bracket of $J_3$ with $E_x$
$$
 \{J_3\,,E_x\} = -\,\frac{2 n_x n_y}{{\lambda}\,\om_0} \Im(M_{xy})
$$
where $\lambda$ is a numerical coefficient.

%-------------------------------------------------------------------
%%  Section 4
\section{Comments and open questions  }

As stated in the introduction the superintegrability of
$V_a(n_x,n_y)$ was firstly proved in \cite{RaSaMontreal} and then
in  \cite{EvVe08b, RodTW08,RodTW09} by the use of different
methods.  Of course these methods are all correct
(the dimensional reduction has been previously applied to the
study of a certain number of integrable systems) but we think that
the approach presented in this paper (deformation of the quadratic
equation) has the great advantage of possessing a great level of
elegance and simplicity.  Moreover it is related with one of the
more  fundamental properties of the harmonic oscillator.

We note that although this method is rather different from the the
dimensional reduction they have in common some important points.
In the geometric method \cite{EvVe08b, RodTW08, RodTW09}, the
authors start with a four dimensional harmonic oscillator  $V_4 =
{1\over2} \sum_{a=1}^4 \omega^2 n_a^2s_a^2$ and then they obtain
the nonlinear system by reducing the dimension from $n=4$ to
$n=2$. In the method presented in this paper we also start with an
harmonic oscillator but then we obtain  the nonlinear system by a
deformation of the complex quadratic equation. Thus, in both cases
the starting point is the (linear) harmonic oscillator, which
is superintegrable in any dimension.

Finally, it is natural to think that the introduction of the
deformation (functions $h_i(x_i)$) in the quadratic equation can
also be applied to equations of order higher  than 2.
Therefore,  a natural generalization of this formalism would be
the search of new superintegrable sytems by introducing
deformations in higher powers $K_i^m$ with $m>2$ of the functions
$K_i$.  Note that we have restricted the study to deformations of
the real part of the functions $K_i^2$; nevertheless  in the more
general case the deformations could  be introduced also in the
imaginary part of the complex functions. This question, as well
as some other related problems (as the properties of the quantum
version of this system) are open questions to be studied.

%----------------------------------------
\section*{Acknowledgments}

We are  indebted to P. Tempesta for valuable comments. MFR
acknowledges support from research projects MTM-2006-10531,
FIS-2006-01225, and DGA-E24/1; MAR from research projects
FIS-2008-00209  and GR-58/08-910556 (Univ. Complutense); and MS
from research project JCyL-GR224-08.

{\small
%---------------

%-------------------------------------------------------------------
\end{document}